\documentclass[letterpaper]{article} % DO NOT CHANGE THIS
\usepackage{aaai2026}  % DO NOT CHANGE THIS
\usepackage{times}  % DO NOT CHANGE THIS
\usepackage{helvet}  % DO NOT CHANGE THIS
\usepackage{courier}  % DO NOT CHANGE THIS
\usepackage[hyphens]{url}  % DO NOT CHANGE THIS
\usepackage{graphicx} % DO NOT CHANGE THIS
\usepackage{natbib}  % DO NOT CHANGE THIS AND DO NOT ADD ANY OPTIONS TO IT
\usepackage{caption} % DO NOT CHANGE THIS AND DO NOT ADD ANY OPTIONS TO IT
\usepackage[table]{xcolor}
\usepackage{pifont}
\newcommand{\Checkmark}{\ding{51}}

\usepackage{algorithm}
\usepackage{algorithmic}

\usepackage{newfloat}
\usepackage{listings}
\DeclareCaptionStyle{ruled}{labelfont=normalfont,labelsep=colon,strut=off} % DO NOT CHANGE THIS
\floatstyle{ruled}
\newfloat{listing}{tb}{lst}{}
\floatname{listing}{Listing}
\usepackage{multirow}
\usepackage{booktabs}

\usepackage{amsmath}
\usepackage{amsfonts}
\usepackage{bm}

\usepackage[hidelinks]{hyperref}
\makeatletter
\let\orig@PackageError\PackageError
\long\def\PackageError#1#2#3{%
  \def\tmp@one{#1}\def\tmp@aaai{aaai}\def\tmp@aaai@{aaai2026}%
  \ifx\tmp@one\tmp@aaai
    \PackageWarning{#1}{#2 (silenced)}%
  \else\ifx\tmp@one\tmp@aaai@
    \PackageWarning{#1}{#2 (silenced)}%
  \else
    \orig@PackageError{#1}{#2}{#3}%
  \fi\fi}
\makeatother
\usepackage{cleveref}  % cleveref MUST be loaded AFTER hyperref
\crefname{section}{Sec.}{Secs.}
\crefname{equation}{Eq.}{Eqs.}
\crefname{figure}{Figure}{Figures}
\crefname{table}{Table}{Tables}
\crefname{algorithm}{Algorithm}{Algorithms}

\title{DiffNR: Diffusion-Enhanced Neural Representation Optimization for Sparse-View 3D Tomographic Reconstruction}
\author{
    Shiyan Su\textsuperscript{\rm 1}\footnote{These authors contributed equally. Project page at \url{https://ooonesevennn.github.io/DiffNR/}}, 
    Ruyi Zha\textsuperscript{\rm 2}\footnotemark[1], 
    Danli Shi\textsuperscript{\rm 3}, 
    Hongdong Li\textsuperscript{\rm 2}, 
    Xuelian Cheng\textsuperscript{\rm 1}\thanks{Corresponding author: Xuelian Cheng}
}
\affiliations{
    \textsuperscript{\rm 1}Monash University\\
    \textsuperscript{\rm 2}The Australian National University\\
    \textsuperscript{\rm 3}Hong Kong Polytechnic University\\
    Xuelian.Cheng@monash.edu
}

\usepackage{bibentry}
\begin{document}
\maketitle

\begin{abstract}
Neural representations (NRs), such as neural fields and 3D Gaussians, effectively model volumetric data in computed tomography (CT) but suffer from severe artifacts under sparse-view settings. To address this, we propose DiffNR, a novel framework that enhances NR optimization with diffusion priors. At its core is SliceFixer, a single-step diffusion model designed to correct artifacts in degraded slices. We integrate specialized conditioning layers into the network and develop tailored data curation strategies to support model finetuning. During reconstruction, SliceFixer periodically generates pseudo-reference volumes, providing auxiliary 3D perceptual supervision to fix underconstrained regions. 
Compared to prior methods that embed CT solvers into time-consuming iterative denoising, our repair-and-augment strategy avoids frequent diffusion model queries, leading to better runtime performance. Extensive experiments show that DiffNR improves PSNR by 3.99 dB on average, generalizes well across domains, and maintains efficient optimization.
\end{abstract}

%\begin{links}
%    \link{Code}{https://aaai.org/example/code}
%    \link{Datasets}{https://aaai.org/example/datasets}
%    \link{Extended version}{https://aaai.org/example/extended-version}
%\end{links}

\section{Introduction}
X-ray computed tomography (CT) is an essential imaging technique for noninvasive inspection of internal structures. A CT scanner captures multi-view projections that record the X-ray attenuation through the material. Given these projections, 3D tomographic reconstruction aims to recover a radiodensity volume. Conventional CT systems acquire hundreds of projections to produce a clean volume, but this results in substantial radiation exposure to subjects. Sparse-view CT (SVCT) reconstruction, which aims to maintain high-quality recovery with only a few dozen projections, thus becomes a crucial direction for safer imaging.

Recent years have seen rapid progress in learning-based SVCT. While feedforward approaches exist~\cite{jin2017deep,lin2024c}, optimization frameworks are generally preferred to enforce consistency between predicted volumes and measured projections. They can be broadly categorized into neural representation (NR) and neural prior (NP) approaches. NR methods model the volume as learnable neural fields~\cite{zha2022naf} or 3D Gaussians~\cite{zha2024r}, and optimize them in a self-supervised manner. They outperform traditional algorithms but yield artifacts in underconstrained regions. In contrast, NP methods pretrain networks to learn data-driven priors and then align network outputs with measurements using optimization solvers. Recent state-of-the-art NP approaches adopts unconditional 2D diffusion models~\cite{ho2020denoising} as network backbone, and embed local solvers into iterative denoising steps. While adequately steering unconditional generation towards the true data manifold, they suffer from inter-slice jitters, hallucinations, and long processing time.

\begin{figure*}[!t]
    \centering
    \includegraphics[width=1.0\linewidth]{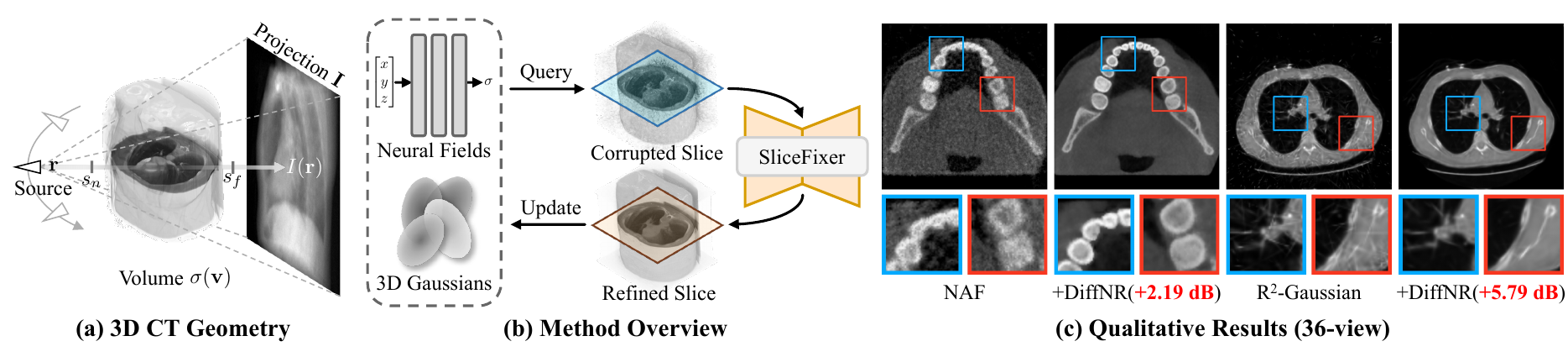}
    \caption{We propose DiffNR for sparse-view 3D CT reconstruction. (a) Geometry of a cone-beam CT scanner. (b) Method overview. (c) Comparison between the baseline methods~\cite{zha2022naf,zha2024r} and our proposed DiffNR.}
    \label{fig:teaser}
\end{figure*}

In this work, we aim to marry neural representations with diffusion models. Unlike prior methods that embed local solvers into unconditional denoising processes, we adopt a fundamentally different strategy: enhancing a global NR with conditioned diffusion models. This design offers clear advantages: (1) learning a unified 3D representation promotes volumetric consistency, and (2) we can finetune powerful 2D foundation models instead of training one from scratch. Nevertheless, this integration is non-trivial, with key challenges in developing an NR-aware diffusion model and efficiently incorporating it into NR optimization.

To tackle these challenges, we propose DiffNR, \textbf{Diff}usion enhanced \textbf{N}eural \textbf{R}epresentation, for sparse-view 3D CT reconstruction. At its core is SliceFixer, a diffusion model specifically adapted to correct artifacts in NR-reconstructed slices. Leveraging 2D foundation models and recent advances in inference acceleration, we finetune a single-step diffusion model~\cite{sauer2024adversarial} on a curated dataset of clean and corrupted slice pairs under varying sparsity levels. To improve structural awareness, we incorporate biplanar X-ray projections as additional conditioning inputs. During the reconstruction phase, SliceFixer periodically generates pseudo-reference volumes, which guide NR optimization in underconstrained regions. We adopt a perceptual SSIM-based regularization instead of voxel-wise losses to mitigate hallucinations and promote structural integrity. This repair-and-augment strategy reduces the need for frequent diffusion model queries, thus ensuring computational efficiency. We evaluate DiffNR across in-distribution and out-of-distribution datasets. Extensive experiments show that it improves NR  reconstruction quality by 3.99 dB, generalizes well across domains, and maintains reasonable runtime.

We summarize our contributions as follows. (1) We propose DiffNR, a novel framework that combines neural representation with diffusion priors, fundamentally different from prior CT methods. (2) We design an effective pipeline to adapt diffusion models for artifact correction and efficiently integrate them into NR optimization, which may also inspire other inverse problems. (3) Experiments demonstrate that DiffNR outperforms existing methods in accuracy, generalization, and efficiency, highlighting its practical values.

\section{Related Work}

\paragraph{Computed Tomography}
CT is widely used in daily applications such as medical diagnosis and security screening. Conventional fan-beam CT reconstructs a 3D volume slice by slice from 1D projection arrays. More recently, cone-beam CT has become popular as it swiftly captures 2D projection images, creating demand for direct volumetric reconstruction. Traditional algorithms fall into direct and iterative methods. Direct approaches~\cite{feldkamp1984practical} instantly compute analytical results but produce severe artifacts. Iterative methods~\cite{andersen1984simultaneous,sidky2008image} formulate reconstruction as an optimization problem and solve it using numerical solvers. They reduce artifacts but oversmooth fine details.

\paragraph{Learning-Based Tomographic Reconstruction}
Similar to traditional algorithms, learning-based CT reconstruction can be performed directly or iteratively. Many works use feedforward networks to predict results from projections~\cite{lin2024c,zhang2025x} or low-quality reconstructions~\cite{jin2017deep,ma2023freeseed}. Such a direct regression, however, lacks physical constraints. Consequently, more attention has shifted to optimization frameworks, broadly grouped into neural representation (NR) and neural prior (NP) approaches. NR methods, inspired by advances in RGB view synthesis such as NeRF~\cite{mildenhall2020nerf} and 3D Gaussian splatting (3DGS)~\cite{kerbl20233d}, optimize a learnable field via differentiable rendering. There are NeRF~\cite{zha2022naf,cai2024structure} and 3DGS~\cite{zha2024r,li20253dgr} variants for 3D CT, but they struggle in sparse-view settings. NP methods combine optimization solvers (traditional or NR-based) with pretrained networks. Some methods use deterministic networks~\cite{kamilov2023plug,tian2025unsupervised,vo2024neural} as regularizer, and the state of the art plugs traditional local solvers into unconditional diffusion models~\cite{chung2023solving,chung2023decomposed}. 
Within this paradigm, there are some early diffusion-NR hybrids~\cite{du2024dper,chu2025highly} which adapt NR as local solvers.
Compared with prior methods, our DiffNR takes a new direction by enhancing a global NR with conditional diffusion models.

\paragraph{Diffusion-Enhanced Neural Representation}
Enhancing NR with diffusion priors has proven to be effective in RGB view synthesis. Some works use diffusion models as scorers that must be queried at each optimization step~\cite{gu2023nerfdiff,warburg2023nerfbusters,zhou2023sparsefusion}, which significantly compromises efficiency. Other approaches finetune diffusion models to repair corrupted images rendered from NR and augment training views with these pseudo-observations~\cite{liu20243dgs,liu2024deceptive}. This strategy avoids frequent diffusion queries, thereby reducing computational overhead. Notably, Difix3D+~\cite{wu2025difix3d+} further improves efficiency by employing single-step diffusion models~\cite{sauer2024adversarial}. Our method follows the repair-and-augment strategy but introduces key innovations designated for CT: (1) we correct artifacts on reconstructed slices rather than on rendered projections, and (2) we augment pseudo-volumes for direct 3D supervision instead of relying on intermediate image losses.

\begin{figure*}[t!]
    \centering
    \includegraphics[width=1.0\linewidth]{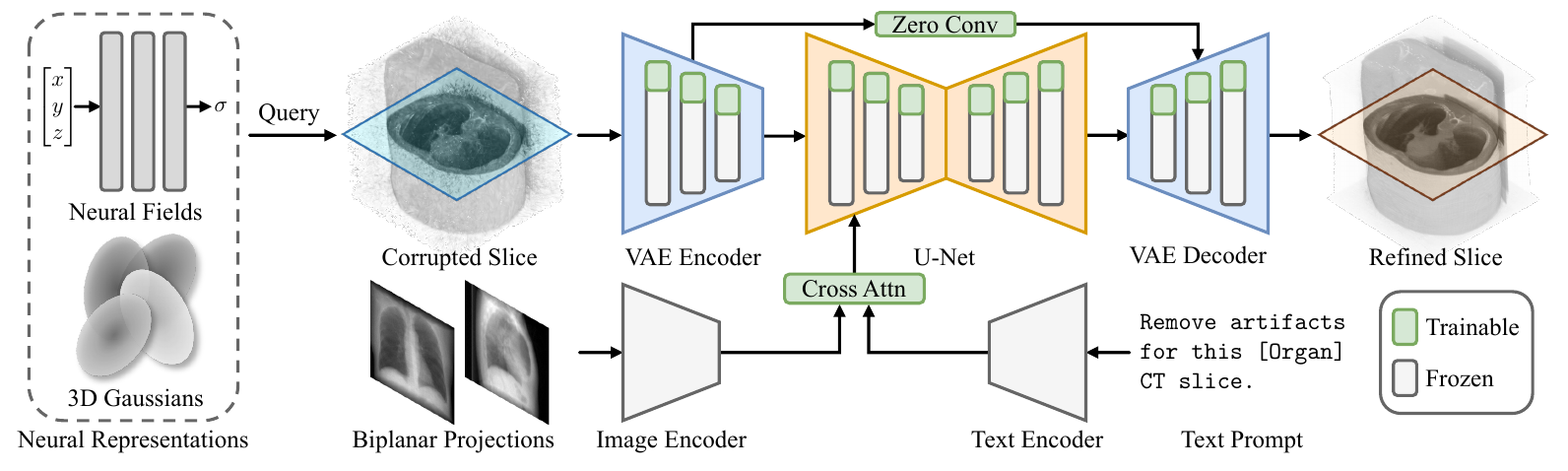}
    \caption{SliceFixer Architecture. It takes as input a CT slice queried from NRs, along with biplanar projections and a text prompt as conditions. It outputs a refined slice without artifacts. The model is built on SD-Turbo~\cite{sauer2024adversarial}, a single-step diffusion backbone. Trainable LoRA layers and zero convolutions are injected to adapt the model for our purpose.}
    \label{fig:diffusion_model}
\end{figure*}

\section{Background}

\paragraph{X-ray Imaging}
%As shown in~\cref{fig:teaser}(a), a projection $\mathbf{I}\in \mathbb{R}^{H\times W}$ measures the attenuation of X-ray beams. 
This work adopts cone-beam geometry as a typical example of 3D CT, and the proposed method can be readily adapted to other geometries such as parallel-beam. As shown in~\cref{fig:teaser}(a), an X-ray with initial intensity $I_0$ travels along the trajectory $\mathbf{r}(s)= \mathbf{o} + s\mathbf{d} \in \mathbb{R}^3$ where $s\in[s_n, s_f]$, passes through a density field $\sigma(\mathbf{v}):\mathbb{R}^3 \to \mathbb{R}$ where $\mathbf{v}$ is any spatial location, and eventually reaches the detector plane. According to the Beer-Lambert law~\cite{kak2001principles}, the corresponding raw pixel value is given by $I'(\mathbf{r})=I_{0}\exp ( -\int^{s_f}_{s_n}\sigma(\mathbf{r}(s))ds)$. In practice, raw data are transformed into logarithmic space for computational convenience, yielding the processed pixel value: $I(\mathbf{r}) = \log I_0 - \log I'(\mathbf{r}) = \int_{s_n}^{s_f}\sigma(\mathbf{r}(s)) \, ds$.
Unless otherwise stated, we use the logarithmic projections as inputs. The goal of tomographic reconstruction is to recover the underlying density field $\sigma(\mathbf{v})$, typically output as a discrete voxel grid $\mathbf{V}\in \mathbb{R}^{X\times Y \times Z}$, from multi-angle projections $\{\mathbf{I}_{i}\}_{i=1}^{N}$. Note that real-world projections contain noise due to physical effects and hardware imperfections.

\paragraph{Neural Representations}
NR methods trains a 3D model via differentiable rendering. There are two primary types of NRs: neural fields and 3D Gaussians. Neural fields, as exemplified by NAF~\cite{zha2022naf}, represent the density field with a multilayer perceptron (MLP) $f$, which can be queried at any location $\mathbf{v}$ to produce the corresponding density $\sigma_{f}(\mathbf{v})$. The rendering function is a discrete Beer-Lambert law: $I_{f}(\mathbf{r}) = \sum_{i=1}^{P} \sigma_{f}(\mathbf{r}(s_{i})) \cdot (\mathbf{r}(s_{i+1}) - \mathbf{r}(s_{i}))$
where \(P\) is the number of sampled points along each ray.

R$^2$-Gaussian~\cite{zha2024r} is a recent 3DGS-based approach, offering faster reconstruction than neural field methods. It represents the density field as a mixture of 3D Gaussians: $\sigma_{g}(\mathbf{v}) = \sum_{i=1}^{M} \mathcal{G}^{3}_{i}(\mathbf{v})$,
where $M$ is the number of kernels. Each Gaussian $\mathcal{G}^{3}_{i}$ has learnable parameters: base density $\rho_i$, center $\mathbf{p}_i \in \mathbb{R}^{3}$, and covariance $\mathbf{\Sigma}_i \in \mathbb{R}^{3\times 3}$. Its form is given by: $\mathcal{G}^{3}_{i}(\mathbf{v}) = \rho_i \exp(-\frac{1}{2} (\mathbf{v} - \mathbf{p}_{i})^{\top} \mathbf{\Sigma}_{i}^{-1} (\mathbf{v} - \mathbf{p}_{i}))$.
To render a projection image, each 3D Gaussian is splatted onto the image plane as a 2D Gaussian $\mathcal{G}^{2}_{i}(\mathbf{u})$, where $\mathbf{u}\in \mathbb{R}^2$. The final projection is then computed by summing all 2D Gaussians: $I_{g}(\mathbf{u}) = \sum_{i=1}^{M} \mathcal{G}^{2}_{i}(\mathbf{u})$. We use NAF and R$^2$-Gaussian as two NR backbones.

\paragraph{Diffusion Models}
Diffusion models~\cite{ho2020denoising,song2020score} learn to approximate the data distribution $p_{\text{data}}$ through iterative denoising. During training, a noisy version of a data sample $\mathbf{x} \sim p_{\text{data}}$ is generated as $\mathbf{x}_{t} = \sqrt{\bar{\alpha}_{t}}\,\mathbf{x} + \sqrt{1 - \bar{\alpha}_{t}}\bm{\epsilon}$, where $\bm{\epsilon} \sim \mathcal{N}(\mathbf{0}, \mathbf{1})$ is standard Gaussian noise, and $\bar{\alpha}_{t}$ controls noise level. The discrete diffusion timestep $t$ is sampled from a uniform distribution $p_t \sim \mathcal{U}(0, t_{max})$. The denoising network $\theta$ predicts the added noise $\bm{\epsilon}_{\theta}$ and is optimized with the score matching objective:
$\mathbb{E}_{\mathbf{x}\sim p_{\text{data}},\, t \sim p_t,\, \bm{\epsilon}\sim \mathcal{N}(\mathbf{0}, \mathbf{1})}\left[\left\lVert \bm{\epsilon} - \bm{\epsilon}_{\theta}(\mathbf{x}_{t}; \mathbf{c}, t)\right\rVert_2^2\right]$,
where $\mathbf{c}$ denotes optional conditioning information, such as text or images. Recent advances~\cite{sauer2024adversarial} accelerate diffusion inference by distilling the multi-step denoising process into a single-step generation.

\section{Proposed Method}

Given $N$ projection images $\{\mathbf{I}_{i}\}_{i=1}^{N}$ acquired at uniform angular intervals around an object, our goal is to reconstruct its volumetric density field $\sigma(\mathbf{v})$, with emphasis on underconstrained regions that are prone to artifacts. To tackle this, we introduce DiffNR, a neural representation optimization framework with diffusion-based augmentation. This section is organized as follows. We begin by introducing SliceFixer, a single-step diffusion model that repairs degraded CT slices. Next, we detail the data curation strategies for model finetuning. Finally, we illustrate how to efficiently integrate SliceFixer into the optimization pipeline.

\subsection{SliceFixer: Diffusion Model for Slice Repairing}

Previous NR methods~\cite{wu2025difix3d+} repair artifacts at the projection level and incorporate intermediate image losses to optimize 3D models. While effective for surface-based RGB reconstruction, this strategy is suboptimal for volumetric reconstruction, where errors in penetrable X-ray projections accumulate. To address this, we propose SliceFixer, a diffusion model that predicts a refined slice $\hat{\mathbf{S}} \in \mathbb{R}^{X' \times Y'}$ from its counterpart $\Tilde{\mathbf{S}}$ queried from NRs. We build SliceFixer upon SD-Turbo~\cite{sauer2024adversarial}, a single-step diffusion model that has demonstrated strong performance in image-to-image translation tasks~\cite{parmar2024one} and providing good inference efficiency. Following \citet{chung2023solving}, we use axial (z-direction) slices in practice, though the approach can be extended to arbitrary slicing directions. Architecture is shown in~\cref{fig:diffusion_model}. A VAE encodes corrupted slices into latents, and a U-Net predicts the target latents conditioned on the encoded inputs, conditions, and denoising timestep. The refined slice is then reconstructed using the VAE decoder.

\paragraph{Conditioning}
% View-Conditioned Diffusion
 We aim to teach SliceFixer to remove artifacts while preserving anatomical structures in CT slices. To this end, our model is conditioned jointly on a text prompt $c_t$ and two orthogonal X-ray projections $(\mathbf{I}_{a}, \mathbf{I}_{b})$. The text prompt provides high-level semantic guidance, whereas the biplanar X-ray projections contains global structural cues. We employ the pretrained RAD-DINO~\cite{perez-garcia_exploring_2025} tailored for radiographs to encode image features. These image features are subsequently aggregated with text embedding via a cross-attention layer to form the conditioning input $\mathbf{c}=\texttt{Embed}(\mathbf{I}_{a}, \mathbf{I}_{b},c_{t})$ for the diffusion model. 

\paragraph{Finetuning}
We finetune a pretrained 2D foundation model SD-Turbo ~\cite{sauer2024adversarial} to leverage its rich visual priors. Following Pix2pix-Turbo~\cite{parmar2024one}, we inject LoRA adapters~\cite{hu2022lora} into the VAE and U-Net modules and incorporate skip connections between the encoder and decoder via zero-convolution layers~\cite{zhang2023adding}. Other parameters are kept frozen.

\paragraph{Losses}
We integrates several standard diffusion losses, including L2 loss, LPIPS loss~\cite{zhang2018unreasonable}, CLIP alignment loss~\cite{radford2021learning}, and an adversarial loss implemented with a CLIP-based discriminator for the target domain~\cite{parmar2024one}. Additionally, we introduce a structural similarity (SSIM)~\cite{wang2004image} loss that captures perceptual quality. Our final objective is defined as:
\begin{equation*}
\begin{aligned}
\mathcal{L}_{\text{total}} =\ 
& \mathcal{L}_{\text{L2}} +
  \mathcal{L}_{\text{LPIPS}} +
  \lambda_{\text{CLIP}} \mathcal{L}_{\text{CLIP}} \\
& +\ \lambda_{\text{GAN}} \mathcal{L}_{\text{GAN}} +
     \lambda_{\text{SSIM}} \mathcal{L}_{\text{SSIM}}.
\end{aligned}
\end{equation*}

\begin{algorithm}[t]
\caption{Diffusion-Enhanced NR Optimization}
\label{alg:diffnr}
\textbf{Input}: Sparse-view projections $\{\mathbf{I}_{i}\}_{i=1}^{N}$, scanner calibration parameters $\{\mathbf{K}_{i}\}_{i=1}^{N}$, neural fields $f$ or 3D Gaussians $g$\\
\textbf{Output}: Density volume $\mathbf{V}$\\
\begin{algorithmic}[1] %[1] enables line numbers
\FOR{$j=1$ to $J$}
\STATE Render projection  $\Tilde{\mathbf{I}}_{i}$ with geometry parameters $\mathbf{K}_{i}$
\STATE Compute L1 and SSIM losses between $\Tilde{\mathbf{I}}_{i}$ and ${\mathbf{I}}_{i}$
\STATE Query volume $\Tilde{\mathbf{V}}_{tv}$ and compute total variation (TV)
\IF{$j \bmod \ell = 0$}
\STATE Query volume $\Tilde{\mathbf{V}}_{\ell}$
\FOR {each axial slice $\Tilde{\mathbf{S}}$ in $\Tilde{\mathbf{V}}_{\ell}$}
\STATE Upsample $\Tilde{\mathbf{S}}$ to match $\texttt{SliceFixer}$ input size
\STATE Generate repaired slice $\hat{\mathbf{S}}$ with $\texttt{SliceFixer}$
\STATE Downsample $\hat{\mathbf{S}}$ back to queried size
\ENDFOR
\STATE Stack repaired slices into a volume $\hat{\mathbf{V}}_{\ell}$
\ENDIF
\IF{$\hat{\mathbf{V}}_{\ell}$ exists and $j \bmod \tau=0$}
\STATE Query $\Tilde{\mathbf{V}}$ and compute its 3D SSIM loss with $\hat{\mathbf{V}}_{\ell}$ 
\ENDIF
\STATE Update $f$ or $g$ based on all losses
\ENDFOR
\STATE Query final volume $\mathbf{V}$ from trained $f$ or $g$
\end{algorithmic}
\end{algorithm}

\subsection{Data Curation}

Training SliceFixer requires a large-scale dataset of paired slices, where one slice contains artifacts typically introduced during NR optimization and the other serves as the clean ground truth. However, no existing dataset satisfies these requirements. To address this, we leverage public 3D CT volumes to synthesize projection data and train a diverse set of neural representations. We explore various strategies to expand the training set and improve data diversity.

\paragraph{View Distribution}
We use the tomography toolbox~\cite{biguri2016tigre} to synthesize $K$ dense projections for each real CT volume over a full $360^\circ$ angular range. To simulate sparse-view scenarios, we randomly sample subsets of these projections to train NR models. We explore both uniformly and non-uniformly distributed view configurations. This variation introduces diverse artifact patterns in the reconstructed volumes, thereby enhancing the model’s robustness to varying sparse-view conditions.

\begin{figure}[t]
    \centering
    \includegraphics[width=1.0\linewidth]{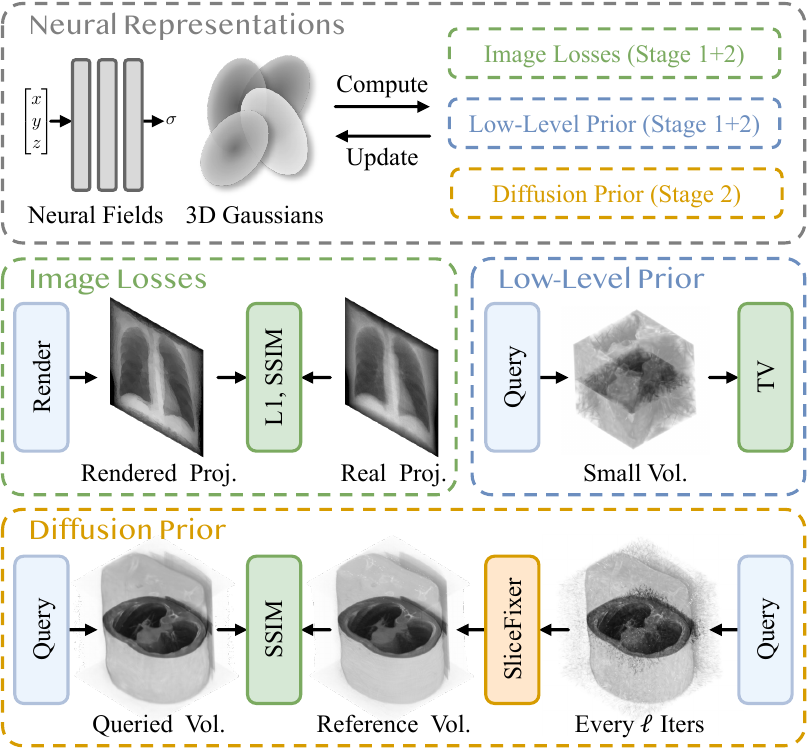}
    \caption{DiffNR Pipeline. During the training, we train neural representations using image losses and low-level regularization. In Stage 2, we generate a 
    pseudo-reference volume with SliceFixer every $\ell$ iterations, and then apply SSIM regularization on queried and reference volumes.}
    \label{fig:pipeline}
\end{figure}

\paragraph{Model Underfitting}
We intentionally underfit the NR optimization by limiting training to a reduced number of iterations (e.g., 25–50\% of the standard training steps). These underfitted reconstructions exhibit more pronounced artifacts due to incomplete convergence, thereby enriching the training set with challenging examples.

\paragraph{Mixed Neural Representation}
We mix reconstruction results from both neural fields and 3D Gaussians in a 1:1 ratio to encourage the diffusion model to learn generalized priors, rather than overfitting to specific patterns.

\begin{table*}[t]  
\centering
\setlength{\tabcolsep}{2.2mm}
\small % 9pt
\begin{tabular}{l|ccc|ccc|c}
\toprule
\multirow{3}{*}{Methods} & \multicolumn{3}{c|}{ToothFairy~\cite{cipriano2022deep}} & \multicolumn{3}{c|}{LUNA16~\cite{setio2017validation}} & \multirow{3}{*}{TIME} \\
\cmidrule(lr){2-4} \cmidrule(lr){5-7}
& 36-view & 24-view & 12-view & 36-view & 24-view & 12-view & \\
& PSNR / SSIM & PSNR / SSIM & PSNR / SSIM & PSNR / SSIM & PSNR / SSIM & PSNR / SSIM & \\
\midrule
\multicolumn{8}{l}{\textbf{Traditional Methods}} \\
SART & 27.41 / 0.581 & 27.13 / 0.596 & 25.66 / 0.604 & 22.34 / 0.438 & 21.77 / 0.437 & 19.96 / 0.412 & 1m25s \\
ASD-POCS & 29.65 / 0.775 & 28.34 / 0.765 & 25.91 / 0.721 & 23.93 / 0.661 & 22.63 / 0.616 & 20.04 / 0.512 & 48s \\
\midrule
\multicolumn{8}{l}{\textbf{Diffusion-Based Iterative Methods}} \\
DiffusionMBIR & \underline{33.29} / 0.856 & 30.54 / 0.818 & 26.28 / 0.733 & \textbf{29.35} / 0.781 & \underline{27.15} / 0.735 & 23.01 / 0.581 & 11h15m \\
DDS & 32.56 / 0.817 & \underline{31.13} / 0.788 & \underline{28.66} / 0.767 & 26.21 / 0.554 & 25.21 / 0.512 & \underline{23.29} / 0.486 & 16m17s \\
\midrule
\multicolumn{8}{l}{\textbf{Neural Representation Methods}} \\
SAX-NeRF & 28.48 / 0.835 & 27.91 / 0.832 & 26.11 / 0.812 & 23.72 / 0.704 & 23.20 / 0.690 & 21.50 / 0.639 & 4h9m \\ 
NAF & 28.62 / 0.833 & 28.20 / 0.833 & 26.22 / 0.812 & 23.85 / 0.712 & 23.18 / 0.692 & 21.37 / 0.618 & 7m15s \\
\rowcolor{gray!20}
\textbf{+DiffNR (Ours)} & 31.27 / \textbf{0.951} & 30.79 / \textbf{0.946} & 28.10 / \textbf{0.906} & 26.27 / \textbf{0.867} & 25.15 / \textbf{0.839} & 22.98 / \textbf{0.765} & 8m41s \\

R$^2$-Gaussian & 28.56 / 0.695 & 26.36 / 0.634 & 22.63 / 0.537 & 24.11 / 0.577 & 22.06 / 0.497 & 18.32 / 0.364 & 5m52s \\
\rowcolor{gray!20}
\textbf{+DiffNR (Ours)} & \textbf{33.52} / \underline{0.900} & \textbf{32.92} / \underline{0.895} & \textbf{29.71} / \underline{0.852} & \underline{28.82} / \underline{0.822} & \textbf{27.43} / \underline{0.793} & \textbf{24.37} / \underline{0.712} & 11min35s \\
\bottomrule
\end{tabular}
\caption{Quantitative results on ToothFairy and LUNA16 datasets. The best values are in bold, second-best are underlined.}
\label{tab:Quantitative_Main}
\end{table*}

\subsection{DiffNR: Diffusion-Enhanced Neural Representation Optimization}
While SliceFixer effectively suppresses artifacts, it may introduce hallucinated details, which is highly undesirable in medical diagnostics. Moreover, this 2D model fails to maintain volumetric consistency, resulting in noticeable inter-slice jitters. To address these issues, instead of treating SliceFixer as a post-processing module, we integrate it into the NR optimization process. DiffNR pipeline is illustrated in ~\cref{fig:pipeline}, and the algorithm is shown in~\cref{alg:diffnr}.
  
\paragraph{Enhanced Volumes as Augmented Supervision}
We begin by optimizing a NR using standard image losses (L1 and SSIM) and low-level 3D regularization (total variation~\cite{rudin1992nonlinear}) to capture global structures. Every $\ell$ iterations, we query a volume $\Tilde{\mathbf{V}}_{\ell}$ from the current model. We then upsample its slices using bilinear interpolation, apply SliceFixer for artifact correction, and downsample the results to the original resolution, producing a pseudo-reference volume $\hat{\mathbf{V}}_{\ell}$. We show in ablation that this up-downsampling strategy improves reconstruction quality. For the remaining training steps, we augment with an additional 3D supervision between the queried volume $\Tilde{\mathbf{V}}$ and this reference volume $\hat{\mathbf{V}}_{\ell}$ every $\tau$ steps. This repair-and-augment strategy reduces the frequency of SliceFixer queries, thus preserving the overall optimization efficiency.

\paragraph{Perceptual Loss for Structural Integrity}
SliceFixer may introduce hallucinated details not perfectly aligned with measured projections. Consequently, directly minimizing voxel-wise L1 loss, as commonly adopted in image supervision, can lead to suboptimal performance. To address this, we adopt a perceptual loss based on 3D SSIM, computed as the average of 2D SSIM scores across axial, sagittal, and coronal planes. This promotes structural coherence and smoothness in underconstrained regions, rather than overfitting to fine-grained, potentially hallucinated details. We use a loss weight $\lambda_{\text{diff}}$ to balance the contribution of 3D SSIM.

\section{Experiments}

\begin{figure*}[h!]
    \centering
    \includegraphics[width=1.0\linewidth]{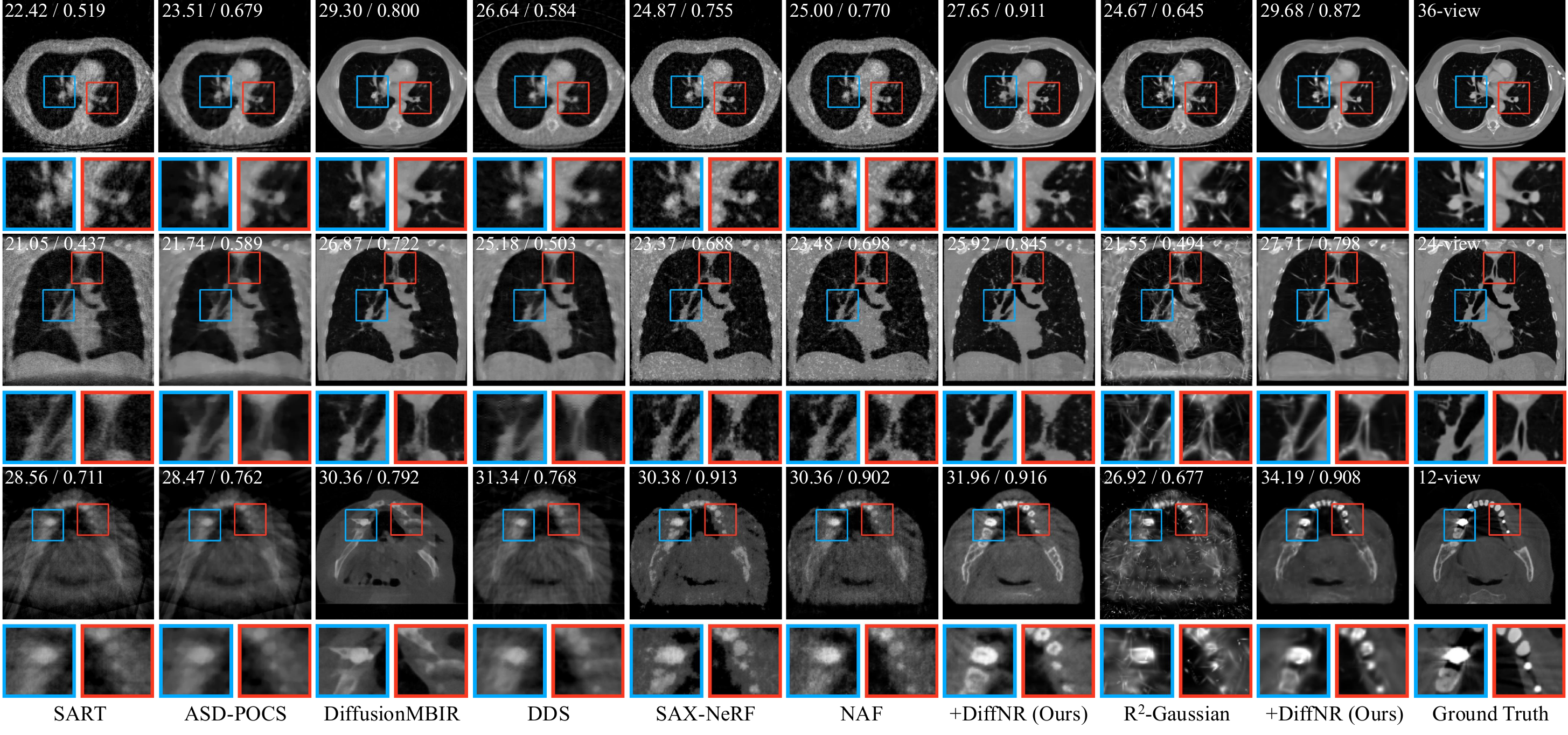}
    \caption{Qualitative results of reconstructed volumes on two datasets, shown from different slicing directions and sparsity levels. We annotate PSNR/SSIM on the top-left of each image. DiffNR recovers finer details and effectively suppresses artifacts.}
    \label{fig:qualitative}
\end{figure*}

\subsection{Experimental Setup}

\paragraph{Datasets}
We use two datasets: ToothFairy~\cite{cipriano2022deep} and LUNA16~\cite{setio2017validation}. ToothFairy consists of 443 dental scans, split into 393/25/25 for training/validation/testing, respectively. LUNA16 includes 888 chest scans, divided into 838/25/25. We train a separate SliceFixer on each dataset and apply the corresponding model for test-case reconstruction. 
We follow~\citet{lin2024c, zha2022naf} to preprocess raw CT volumes to a resolution of $256^3$ and X-ray projections to $256^2$. Sparse-view reconstruction is defined as using fewer than a hundred views, and we evaluate the challenging 36-, 24-, and 12-view settings. %More dataset details are described in the appendix. 
 
\paragraph{Implementation Details}
SliceFixer is finetuned from SD-Turbo on $512^2$ images, which are upsampled from $256^2$ slices. We integrate LoRA layers with ranks of 8 for the U-Net and 4 for the VAE, and train the model with a learning rate of $1\mathrm{e}{-5}$ for 40k steps on ToothFairy and 70k steps on LUNA16, using a batch size of 4. Loss weights are set to $\lambda_{\text{CLIP}} = 4$, $\lambda_{\text{GAN}} = 0.4$, and $\lambda_{\text{SSIM}} = 0.5$. Finetuning is performed on 4 H100 GPUs. DiffNR is implemented in PyTorch and optimized using the Adam optimizer~\cite{kingma2014adam}. We use NAF and R$^2$-Gaussian as backbones, training them for 11k and 13.5k epochs, respectively, while keeping other hyperparameters unchanged. We empirically set $\ell = 10\mathrm{k}$, and use $\tau = 20$ for NAF and $\tau = 10$ for R$^2$-Gaussian. Pseudo-reference volumes have a resolution of $256^3$. All test-case reconstructions are performed on an RTX 3090 GPU. The code and model will be publicly available.

\paragraph{Compared Methods and Evaluation}
We compare with widely-used optimization-based methods, including (1) traditional iterative methods SART~\cite{andersen1984simultaneous} and ASD-POCS~\cite{sidky2008image}, (2) self-supervised NR methods SAX-NeRF~\cite{cai2024structure}, NAF~\cite{zha2022naf}, and R$^2$-Gaussian~\cite{zha2024r}, and (3) diffusion-based iterative methods: DDS~\cite{chung2023decomposed} and DiffusionMBIR~\cite{chung2023solving}. We quantitatively evaluate all methods using standard metrics PSNR and SSIM. %See appendix for details of metrics, implementations, and comparisons with feedforward methods.

\subsection{Results}

\begin{table}[t]
\centering
\small
\setlength{\tabcolsep}{1.1mm}
\begin{tabular}{l|ccc}
\toprule
& \multicolumn{3}{c}{OOD Dataset~\cite{zha2024r}} \\
\cmidrule(lr){2-4}
Methods & 36-view & 24-view & 12-view \\
& PSNR / SSIM & PSNR / SSIM & PSNR / SSIM \\
\midrule
%FDK      & 19.51 / 0.136 & 17.45 / 0.094 & 14.24 / 0.051 \\
SART   & 30.50 / 0.740 & 29.43 / 0.721 & 27.53 / 0.695 \\
ASD-POCS        & 32.28 / 0.852 & 30.16 / 0.811 & 27.36 / 0.750 \\
\midrule
DiffusionMBIR  & 33.26 / 0.839 & 30.97 / 0.796 & 26.82 / 0.668 \\
DDS        & 29.45 / 0.638 & 26.97 / 0.536 & 25.17 / 0.520 \\
\midrule
R$^2$-Gaussian         & \underline{35.64} / \underline{0.904} & \underline{33.46} / \underline{0.868} & \underline{29.71} / \underline{0.792} \\
\rowcolor{gray!20}
\textbf{+DiffNR (Ours)}               & \textbf{35.99} / \textbf{0.918} & \textbf{34.15} / \textbf{0.896} & \textbf{31.04} / \textbf{0.848} \\
\bottomrule
\end{tabular}
\caption{Quantitative results on the OOD dataset. The best values are in bold, second-best are underlined} 
\label{tab:Quantitative_OOD}
\end{table}

\paragraph{In-Distribution Performance}
Table~\ref{tab:Quantitative_Main} presents quantitative results on ToothFairy and LUNA16. Traditional methods and self-supervised NR approaches produce significant artifacts. While diffusion-based methods achieve higher scores, they come at the cost of hallucinated details and  significant computation time. Previous SOTA DiffusionMBIR takes 11 hours to process a single case. In contrast, our DiffNR consistently enhances NR baselines, yielding an average improvement of +2.19 dB in PSNR for NAF and +5.79 dB for R$^2$-Gaussian. Although DiffNR introduces additional optimization time, it remains substantially faster than prior diffusion-based methods. Qualitative comparisons are provided in~\cref{fig:qualitative}, where DiffNR recovers fine structures and substantially reduces artifacts present in NR baselines.

\begin{figure}[t]
    \centering
    \includegraphics[width=1.0\linewidth]{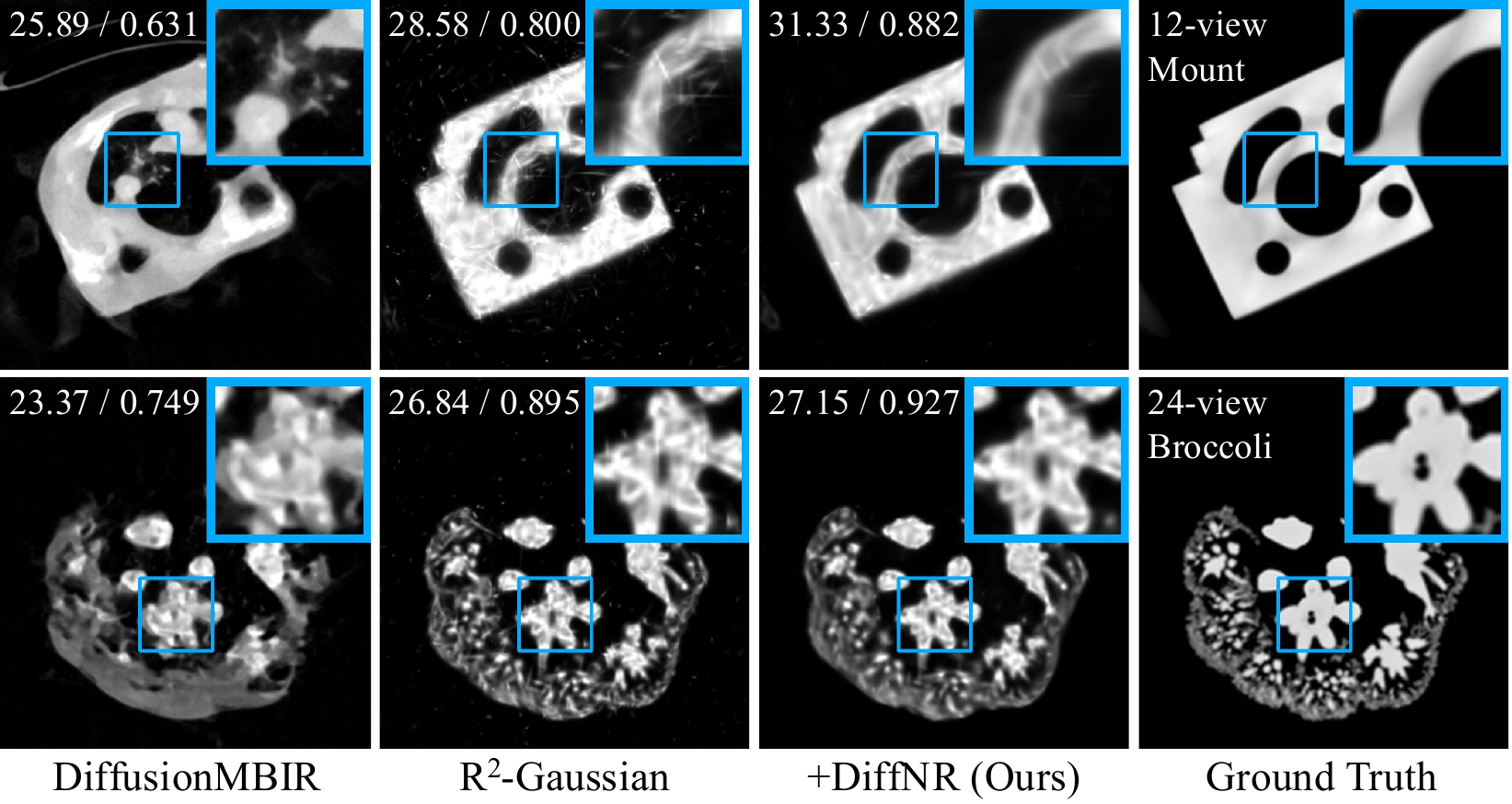}
    \caption{Qualitative results on OOD dataset.}
    \label{fig:ood}
\end{figure}

\paragraph{Out-of-Distribution Performance}
To evaluate generalization capability, we use SliceFixer pretrained on ToothFairy and apply R$^2$-Gaussian+DiffNR to dataset from~\citet{zha2024r}, which includes 18 diverse cases spanning human organs, biological specimens, and artificial objects. Notably, this dataset contains real-world captured projections. Quantitative and qualitative results are shown in~\cref{tab:Quantitative_OOD} and~\cref{fig:ood}, respectively. DiffNR outperforms other methods by suppressing hallucinations and artifacts, which shows that SliceFixer learns generalizable artifact patterns.

\begin{figure}[t]
    \centering
    \includegraphics[width=1.0\linewidth]{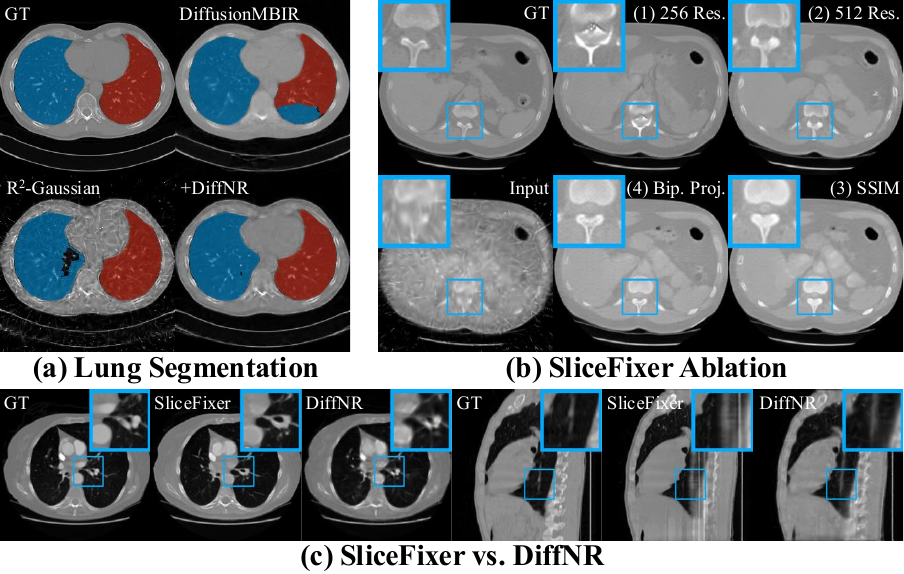}
   \caption{Qualitative results of downstream tasks and ablation study. (a) Lung segmentation with the left lung in blue and the right lung in red. (b) Visualization of different design choices for SliceFixer. (c) Comparison of standalone SliceFixer post-processing and our proposed DiffNR.}
   \label{fig:seg_and_slicefixer}
\end{figure}

\paragraph{Downstream Application}
We further validate our method on downstream medical tasks such as segmentation. Specifically, we use the LungMask toolkit~\cite{hofmanninger2020automatic} to perform left/right lung segmentation on the reconstructed volumes. We use Dice~\cite{dice1945measures} and average surface distance (ASD) metrics to evaluate performance. As shown in~\cref{tab:segmentation} and~\cref{fig:seg_and_slicefixer}(a), the segmentation masks generated from Gaussian-based DiffNR are more consistent with those obtained from the ground-truth volumes, demonstrating the practical utility of our method.

\begin{table}[t]
\centering
\small
\setlength{\tabcolsep}{1.4mm}
\begin{tabular}{l|ccc}
\toprule
\multirow{2}{*}{Methods} & 36-view & 24-view & 12-view \\
                         & Dice$\uparrow$/ASD$\downarrow$ & Dice$\uparrow$/ASD$\downarrow$ & Dice$\uparrow$/ASD$\downarrow$ \\
\midrule
%FDK      & 72.73 / 23.26 & 69.03 / 24.80 & 54.68 / 29.32 \\
SART     & 81.89 / 11.73 & 74.12 / 16.63 & 56.92 / 26.62 \\
ASD-POCS & 76.47 / 15.71 & 70.06 / 19.64 & 57.98 / 22.27 \\
\midrule
DiffusionMBIR  & 90.33 / 6.13 & \underline{86.96} / \underline{6.97} & \underline{77.75} / \underline{11.96} \\
DDS            & 80.03 / 16.66 & 75.98 / 16.60 & 68.23 / 19.20 \\
\midrule
R$^2$-Gaussian & \underline{90.41} / \underline{5.19} & 84.32 / 8.39 & 59.73 / 25.11 \\
\rowcolor{gray!20}
\textbf{+DiffNR (Ours)} & \textbf{93.74} / \textbf{3.85} & \textbf{90.71} / \textbf{5.60} & \textbf{84.93} / \textbf{9.59} \\
\bottomrule
\end{tabular}
\caption{Quantitative results for lung segmentation of reconstructed results on LUNA16 dataset.}
\label{tab:segmentation}
\end{table}

\begin{table}[t]
\centering
\small
\setlength{\tabcolsep}{1mm}
\begin{tabular}{c cccc|c|c}
\toprule
ID & Res. & SD-Turbo Pretrain & \(\mathcal{L}_{ssim}\) & Bip. Proj. & PSNR & SSIM  \\ \midrule
(1) & 256   & \Checkmark      &    &  & 27.65 & 0.789 \\
(2) & 512    & \Checkmark    &   &   & 27.91 & 0.807 \\
(3) & 512 & \Checkmark & \Checkmark    &      & 28.21 & 0.814 \\
(4) & 512 & \Checkmark  & \Checkmark   & \Checkmark  & \textbf{28.82}& \textbf{0.822}\\ 
\bottomrule
\end{tabular}
\caption{Ablation study of SliceFixer. We finetune different models and evaluate DiffNR under LUNA16 36-view cases.}
\label{tab:ablation_slicefixer_design}
\end{table}

\subsection{Ablation Study}
\paragraph{SliceFixer Design}
We validate design choices of SliceFixer in~\cref{tab:ablation_slicefixer_design} and~\cref{fig:seg_and_slicefixer}(b). We find that finetuning SliceFixer on $512^2$ images and applying up-downsampling to queried slices leads to better reconstruction quality compared to using the original $256^2$ resolution. Additionally, incorporating an SSIM loss into finetuning resulting in a 0.3 dB gain in PSNR. Finally, adding biplanar projections as additional conditioning inputs provides rich structural cues and further boosts finetuning performance by 0.6 dB in PSNR.

\paragraph{DiffNR Design}
We use R$^2$-Gaussian as backbone to validate components of DiffNR as shown in~\cref{tab:ablation_diffnr_design}. Augmenting NRs with novel-view images, commonly used in RGB surface reconstruction~\cite{wu2025difix3d+}, is ineffective in volume reconstruction. This is because errors in penetrable projections can accumulate to the target volume across views. Instead, we choose to augment slice supervision, which proves to be more stable and effective. Moreover, applying SliceFixer as a standalone post-processing step leads to slice jitter and hallucinations (\cref{fig:seg_and_slicefixer}(c)), highlighting the necessity of integrating it into the optimization pipeline. Lastly, we find that using voxel-wise L1 loss results in a performance drop, as the pseudo-reference volumes may contain details inconsistent with measured projections. A 3D perceptual loss is thus preferred. Overall, integrating our proposed components leads to the best performance.

\begin{table}[t]
\centering
\small
\setlength{\tabcolsep}{4.3mm}
\begin{tabular}{l|c|c}
\toprule
Methods & PSNR & SSIM \\
\midrule
R$^2$-Gaussian & 24.11 & 0.577 \\
+ Difix3D+ (augment projection) & 23.23 & 0.579 \\
+ SliceFixer (post-processing) & 26.70 & 0.776 \\
+ SliceFixer (\(\mathcal{L}_{1}\)) & 26.42 & 0.678 \\
+ SliceFixer (\(\mathcal{L}_{ssim}\)) (Ours) & \textbf{28.82} & \textbf{0.822} \\
\bottomrule
\end{tabular}
\caption{Ablation study of DiffNR design on LUNA16 dataset under 36-view setting.}
\label{tab:ablation_diffnr_design}
\end{table}

\begin{table}[!h]
\small
\centering
\setlength{\tabcolsep}{2.4mm}{
\begin{tabular}{ccccccc}
\toprule
$\lambda_{\text{diff}}$ & 0.3 & \textbf{0.5} & 0.7 & 1.0 & 1.5 \\
\midrule
PSNR & 28.65 & 28.82 & 28.79 & 28.72 & 28.63 \\
\midrule
$\tau$ & 5 & \textbf{10} & 15 & 20 & 30 \\
\midrule
PSNR & 28.76 & 28.82 & 28.67 & 28.43 & 27.87 \\
TIME & 27m35s & 12m56s & 10m02s & 8m32s & 7m26s \\
\bottomrule
\end{tabular}}
\caption{Ablation study of DiffNR hyperparameters on LUNA16 dataset (36-view) with our choices in bold.}
\label{tab:ablation_parameter_analysis}
\end{table}

\paragraph{Parameter Analysis}
We perform parameter analysis for Gaussian-based DiffNR to investigate the impact of 3D SSIM loss weight $\lambda_{\text{diff}}$ and 3D supervision frequency $\tau$. As shown in~\cref{tab:ablation_parameter_analysis}, $\lambda_{\text{diff}} = 0.5$ achieves the best performance by balancing the guidance from 3D supervision and avoiding overfitting to projections or degradation from diffusion hallucination. For the supervision interval, $\tau=10$ yields optimal results. More frequent supervision (e.g., $\tau = 5$) may lead to over-reliance on the 3D loss and increased computational cost, whereas sparse supervision (e.g., $\tau = 20$) weakens structural regularization and degrades performance.

\section{Conclusion}
We present DiffNR, a novel optimization framework for sparse-view 3D tomographic reconstruction. At its core is SliceFixer, a single-step diffusion model finetuned on curated datasets to correct artifacts in reconstructed CT slices. During reconstruction, the pretrained SliceFixer generates pseudo-reference volumes that provide augmented perceptual regularization. Such a repair-and-augment strategy avoids frequent diffusion model queries, therefore improving reconstruction quality without sacrificing efficiency. Experimental results demonstrate that DiffNR outperforms prior methods in reconstruction quality, generalization capability, and optimization efficiency, highlighting its practical potential. Further, this novel integration of diffusion models with neural representation optimization opens a promising direction for addressing broader classes of inverse problems.

\section{Acknowledgments}
This research is supported in part by the Jiangsu Department of Technology Natural Science Fund (Grants No: BK20250441), the Center of Excellence for Antimicrobial Therapeutics Discovery and Innovation (CEATDI, Grants No: 8002003), and the ARC Discovery Grant (Grant ID: DP220100800) of the Australia Research Council.

\bibliography{main}

\end{document}